\documentclass[aps,showpacs,pre,twocolumn,amsfonts,floats]{revtex4}

\usepackage{graphicx}
\begin{document}

\title{Scale-Free topologies and Activatory-Inhibitory interactions}

\author{J. G\'omez-Garde\~nes$ ^{(a),(c),(d)}$, Y. Moreno$^{(b),(c)}$ and
L.M. Flor\'{\i}a$^{(a),(c),(d)}$}

\affiliation{$^{(a)}$ Dept.\ de F\'{\i}sica de la Materia Condensada,
  University of Zaragoza, 50009~Zaragoza, Spain}
\affiliation{$^{(b)}$ Dpt. F\'{\i}sica Te\'orica, University of Zaragoza,
  50009~Zaragoza (Spain).}
\affiliation{$^{(c)}$ Instituto de Biocomputaci\'on y F\'{\i}sica de Sistemas
  Complejos BIFI, University of Zaragoza, 50009~Zaragoza, Spain}
\affiliation{$^{(d)}$ Dpto.\ de Teor\'{\i}a y Simulaci\'on de Sistemas
  Complejos.
  Instituto de Ciencia de Materiales de Arag\'on ICMA,
  C.S.I.C. - Universidad de Zaragoza, 50009~Zaragoza, Spain}

\date{\today}

\begin{abstract}
A simple model of activatory-inhibitory interactions controlling the
activity of agents (substrates) through a ``saturated response''
dynamical rule in a scale-free network is thoroughly studied. After
discussing the most remarkable dynamical features of the model, namely
fragmentation and multistability, we present a characterization of the
temporal (periodic and chaotic) fluctuations of the quasi-stasis
asymptotic states of network activity. The double (both structural and
dynamical) source of entangled complexity of the system temporal
fluctuations, as an important partial aspect of the Correlation {\em
Structure-Function} problem, is further discussed to the light of the
numerical results, with a view on potential applications of these
general results.
\end{abstract}
\maketitle

{\bf Many real networks are complex and heterogeneous. In this paper,
  we study a dynamics that generically describes biological processes
  that take place on complex architectures as metabolic reactions and
  gene expression. We capitalize on the theory of nonlinear dynamical
  systems to uncover the topological and dynamical patterns of a
  Michaelis-Menten like dynamics coupled to a network that is complex,
  directed and highly skewed. The results indicate that such patterns
  can exist in the form of periodic and chaotic orbits revealing
  interesting properties at both local and global levels. Moreover, 
  the dynamics on top of the substrate networks yields topologically 
  complex substructures (islands or clusters) whose structural 
  charateristics are analysed. This analysis offers interesting results 
  on the interplay {\em Structure-Function}. We round off our study 
  by discussing possible implications of heterogeneous topologies on 
  biological processes at the cellular level.}

\section{Introduction}
\label{sec:1}

Nonlinear lattices, {\em i.e.} spatially discrete many-body systems
with nonlinear interactions, are currently the subject of a considerable
multidisciplinary interest, not only among a wide variety of Physics
subdisciplines and technologies \cite{scott}, but also in Biomolecular
Chemistry, Cell physiology, Theoretical Biology, Social Sciences and
other fields \cite{introd}. There is a common basic interest in the
understanding of the many aspects of the correlation between
{\em Structure} and {\em Function} in systems made up of discretely
many nonlinearly interacting components.

While in Physics applications the interactions among constituents (atoms,
magnetic moments, {\ldots}) usually depends on the geometrical distance
between their positions in real space, in applications outside ``fundamental
physics'' the space of interactions is abstracted, so that proximity between
components (say $i$ and $j$) is measured as length of the path of
interactions given by a connectivity matrix $C_{ij}$. In other words, the
graph (or network) of interactions, and not anymore the real space, fixes the
relevant geometry related to the function (dynamics) of the system. This
does not preclude the applicability of Statistical and Field theory methods
to the study of nonlinear lattices outside the traditional physics
subdisciplines; on the contrary, a proper use of these approaches often throws
considerable light on some important issues currently addressed
\cite{book1}.

Though lattice disorder effects on nonlinear dynamics of
macroscopic systems have their own tradition, the most usual case in
physics is that of homogeneous (either pure random or regular)
networks. However, recent confluent studies on the {\em Structure}
of interactions in a large variety of technological
(communication, power grid) as well as biomolecular
(protein-protein interaction, gene regulation, cell metabolism),
ecological (trophic networks, mutualism) and socio-economic
systems have shown the overabundance of highly inhomogeneous
structures \cite{RMP,strogatz} among ``real world'' interacting
systems.

Homogeneity of the interactions structure means that almost all nodes are
topologically equivalent, like in a regular lattice or in a random
Erd\"{o}s-Renyi graph, thus showing a density distribution function of the
degree of connectivity localized around a mean value with a well-defined
average of quadratic fluctuations. If $k$ denotes the degree (number of
interactions of a given node) and $P(k)$ denotes its density distribution
function, an inhomogeneous network shows for $P(k)$ a power law (often
truncated). The absence of a characteristic scale in the
connectivity patterns (scale-free networks) manifests itself in the
presence of a few number of nodes (named hubs) connected to very many
nodes, and a larger number of poorly connected nodes. The
complex character of the structure of the interactions 
couples to the dynamical complexity which emerges from the nonlinear
character of the interactions, so that generally one may say that
the {\em Structure-Function} correlation problem in real networks has at
least two sources of entangled complexity.

The model that we analyze is introduced in section \ref{sec:2}. It
tries to capture general ingredients of this entangled complexity in a
relevant kind of nonlinear dynamics: Activation/Inhibition (AI)
competing interactions with a ``saturated response'' rule for the rate
of activation (see Figure 1). This kind of dynamics is often called
Michaelis-Menten \cite{segel}, Holling \cite{drossel}, or Langmuir
\cite{diu} rule. The interacting units sit on a lattice which is both
{\em small-world} ({\em i.e.}  short mean path length) and {\em
scale-free}. For this we use the Barab\'asi- Albert \cite{bar99}
network. Afterwords, some basic general features of the model are
discussed, namely the network fragmentation in subclusters (or
islands) of collective dynamics (\ref{fragmentation}), and the generic
types of asymptotic behaviours coexisting in the phase space of
collective dynamics as well as the observed bifurcations in phase
portrait upon parameter variations (\ref{subsec:2.2}). These basic
consequences of the AI competition on the complex network are
prevalent for values of the ratio AI ranging from 1 to 6.  Finally, in
section \ref{subsec:2.3} the bifurcations found are explained in terms
of the Floquet analysis of the solutions.

Section \ref{sec:3} mainly reports on the statistical characterization
of both the dynamical behaviours observed (section \ref{subsec:3.1})
and the structural characterization of the dynamical islands (section
\ref{subsec:3.2}). We perform an extensive exploration of the
parameter space, employing different initial conditions and substrate
network realizations, in order to find the conditions for the
existence of chaotic and periodic behavior as well as to fully
characterize the main topological characteristics of the dynamical
islands.  Finally, in section \ref{subsec:3.3} we identify those
substructures of the dynamical islands that are relevant for the
dynamical evolution of the system.

The concluding section \ref{sec:4} summarizes the main conclusions of
our work, along with some prospective remarks on likely applications
of the model, and the potential use of these techniques in the study
of particularly interesting real-world biological networks.

\section{The model. Basic dynamical features.}
\label{sec:2}

As stated in the introductory section \ref{sec:1}, we introduce here a model of
Activatory/Inhibitory interactions regulating the activity $g_i(t)$
($i= 1, .., N$), of $N$ constituents ({\em e.g.} agents, substrates), with
$N$ generally being a large number. The real functions of time $g_i(t)$
are each one attached to a node of a graph with adjacency matrix $C_{ij}$
($N \times N$). The matrix element is non-zero, $C_{ij} \neq 0$, only
if the rate of variation of the $i$-th node activity, $g_i(t)$, depends on the
activity $g_j$ of the $j$-th node (interaction $i \leftarrow j$). Different
realizations of the $C_{ij}$ matrix are constructed using the method of
Barab\'asi and Albert \cite{bar99}, in order to ensure two seemingly
universal characteristics of many recently studied networks in Biological and
Social sciences and other fields \cite{RMP,strogatz}:

\begin{itemize}

\item[(a)] {\em Small-world}, meaning that the mean distance
(minimal length of the interaction path), $\langle l_{ij}
\rangle$, between pairs of nodes goes at most as $\log N$, for
large values of $N$.

\item[(b)] {\em Scale-free}, meaning that the density distribution function $P(k)$
of the degree (connectivity) of nodes scales as $P(k) \sim k^{-\gamma}$,
with $\gamma = 3$. Other values of $\gamma$ ($2\leq \gamma \leq 3$) were
also analyzed by using suitably tested modifications of the Barab\'asi-Albert
preferential attachment rule \cite{gamma}.

\end{itemize}

The interaction ($i \leftarrow j$) can be either activatory
(excitatory) or inhibitory; correspondingly we define the interaction
matrix element $W_{ij}$ to be $+1$ or $-1$, respectively (and
$W_{ij}=0$ whenever $C_{ij}=0$), and call $p$ the fraction, among
non-zero elements, of negative signs (note that while $C_{ij}$ is a
symmetric matrix, $W_{ij}$ is not in general). Moreover, the sign
distribution of elements is taken uniform in the set of
(approx. $\langle k \rangle N/2$) links of the network realization.

\begin{figure}[!thb]
\begin{tabular}{cc}
\centerline{
\resizebox{7.5cm}{!}{%
\includegraphics[angle=-0]{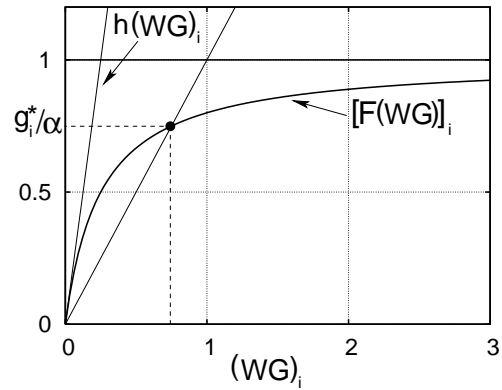}
}
}
\end{tabular}
\caption{Saturated functional response ($h=4$).}
\label{figure1}
\end{figure}

\begin{figure*}[!tb]
\begin{tabular}{cc}
\centerline{
\resizebox{16.cm}{!}{%
\includegraphics[angle=-0]{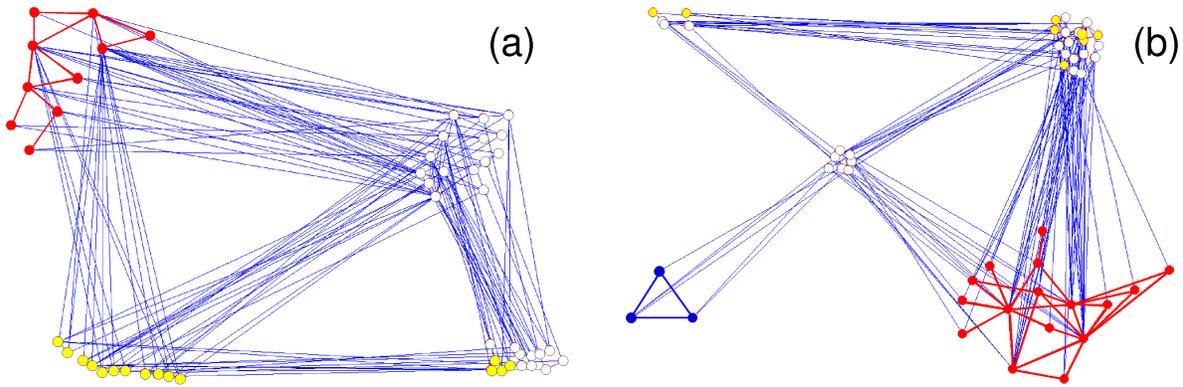}
}
}
\end{tabular}
\caption{(Color online) Two examples of network fragmentation. The nodes 
of the networks are clasified in: ({\it i}) dynamical nodes (red), 
({\it ii}) stationary nodes with nonzero activity (blue), 
({\it iii}) stationary nodes with zero activity beloging to 
${\cal D}_{0}\oplus{\cal D}_{1}$ (yellow) and 
({\it iv}) remaining nodes with zero activity (white). 
Note that the white central nodes in {\bf (b)} act as the 
frontier between the dynamical island and the steady nonzero activity one.}

\label{figure2}
\end{figure*}

The dynamics of the nodes activity vector ${\bf G}(t)=\{g_i(t)\}$ 
(with $i= 1,{\ldots}, N$) that we consider is such that only in the pressence of
excitatory neighbours activity could $g_i$ possibly be non null, otherwise
$g_i$ decays to zero with an exponential rate:
\begin{equation}
\frac{d{\bf G}(t)}{dt}=-{\bf G}(t)+\alpha {\bf F}({\bf
WG}(t)), \label{eq1}
\end{equation}
where ${\bf F}$ is a nonlinear vector function whose argument
is the product of the interaction matrix ${\bf W}$ and the
activity vector ${\bf G}$, and $\alpha$ ($>0$) accounts for the
strength of the interaction. The function ${\bf F}$ is a
saturated response function (see figure \ref{figure1}), defined
as:

\begin{equation}
{\bf F}({\bf z}) = \left\{\frac{\Phi(z_i)}{h^{-1} +\Phi(z_i)}\right\},
\label{saturated}
\end{equation}
where $\Phi(x)$ is a function defined as $\Phi(x)=x$ if $x\ge 0$
and zero otherwise. In our numerical studies of the model we have
fixed the value of the parameter $\alpha = 3$, and varied the
parameters $0 \leq p \leq 1 $ and $0 \leq h \leq 10$. One can
easily realize that the solutions for non-negative initial
conditions remain bounded for all $t \geq 0$. Indeed, as the
nonlinear term in (\ref{eq1}) is bounded above by $\alpha$, one
obtains that $\dot{g}_i < 0$ whenever $g_i > \alpha$. Also, if
$g_i =0$ then $F_i({\bf W}{\bf G}) \geq 0$, so that the
activities cannot become negative.

The above dynamics can be regarded, {\em e. g.}, as a generalization
of some simplified and coarse-grained genetic models, referred to as
Random Boolean Networks \cite{kauffman}, where Boolean rules are
implemented. These have been extensively used to study networks
at various levels of biological organization, and have provided
useful insights later supported by experiments \cite{kauffman,oosawa}.
Equation (\ref{eq1}) incorporates the experimental observation of a
continuous range of activity levels \cite{sole}. While linear models
have been successful for the reconstruction of the interaction networks
from experimental data \cite{reconst}, nonlinear models like
Eqs.\ (\ref{eq1}) are expected to be more appropriate for a
quantitative description of the dynamics.

The dynamics (\ref{eq1}) of a two-agent (dimer) model has been considered
in reference \cite{sole}, in the context of virus-cell interactions in
bacteria and general gene regulatory activity models, where a rich
repertoire of behaviours, like multi-stability (multiple attractors in
phase space) was reported. A preliminary study of the behaviour of
(\ref{eq1}) on small-world scale-free networks can be found in \cite{JGG},
where the interested reader can find a more detailed account of the
numerical techniques used in the characterization of the different
dynamical regimes. In this section, we review some remarkable general
features of the network dynamics.

\subsection{Activation and Inhibition interplay. Fragmentation.}
\label{fragmentation}

A brief look at equation (\ref{eq1}) easily reveals that for any
value of the parameters $p$ and $h$ the state of inactivity, ${\bf
G}=0$, is always a solution. As a matter of fact, for $h= 0$, or
$h \neq 0$ but $p = 1$, this is the unique asymptotic solution
(global attractor in the phase space) for all possible
non-negative initial conditions. However, for $h \neq 0$ and $p
\neq 1$ other asymptotic solutions, with islands of positive
activity, generically coexist with the rest state. The term
islands denotes here subnetworks that are interconnected through
nodes which have evolved to null activity, so that their dynamics
are effectively disconnected.

The fragmentation of the network dynamics into disconnected islands is
a generic feature of AI interactions, as the following considerations
suggest. Let us call ${\mathcal D}$ the set of nodes whose activities,
for a given initial condition ${\bf G}(t=0)$, asymptotically
vanish. It is easy to see that, irrespective of the initial condition,
this set is generically non-empty.

Indeed, if a node $i$ is such that $W_{ij} =-1$ or $0$ for all $j$,
then its activity $g_i(t)$ will tend to zero. Let us call ${\mathcal
D}_0$ the set of these nodes, and note that its measure ($\sum_k P(k)
p^k$) is a non-zero increasing function of $p$. Now, call ${\mathcal
D}_1$ the set of nodes $l$ such that their positive $W_{lj}$ occur for
$j$'s in ${\mathcal D}_0$, and so on {\ldots} Due to the small-world
property, there are in fact very few relevant ${\mathcal D}_n$
($n=0,1,{\ldots}$) sets. Its union ${\mathcal D}^{*} = \bigcup
{\mathcal D}_n$ is easily seen to have a non-zero measure which
increases with $p$.

The nodes of ${\mathcal D}^{*}$ are structurally ({\em i.e.} irrespective
of initial conditions) inactive. Depending on the initial condition, the
set ${\mathcal D}$ may include other nodes not contained in ${\mathcal D}^{*}$,
namely those nodes that evolve to inactivity due to the initial condition
(dynamically, instead of structurally, inactive): See {\em e.g.} the
white nodes in figure \ref{figure2}, where we show two small networks of $N=50$
nodes to allow a simple visualization of the sets ${\mathcal D}^{*}$ and
${\mathcal D}$. In other words, the measure of ${\mathcal D}$ may
in general be (much) larger than the measure of the ``structurally
dead'' nodes ${\mathcal D}^{*}$.

\begin{figure*}[!htb]
\begin{tabular}{cc}
\centerline{
\resizebox{14.cm}{!}{%
\includegraphics[angle=-0]{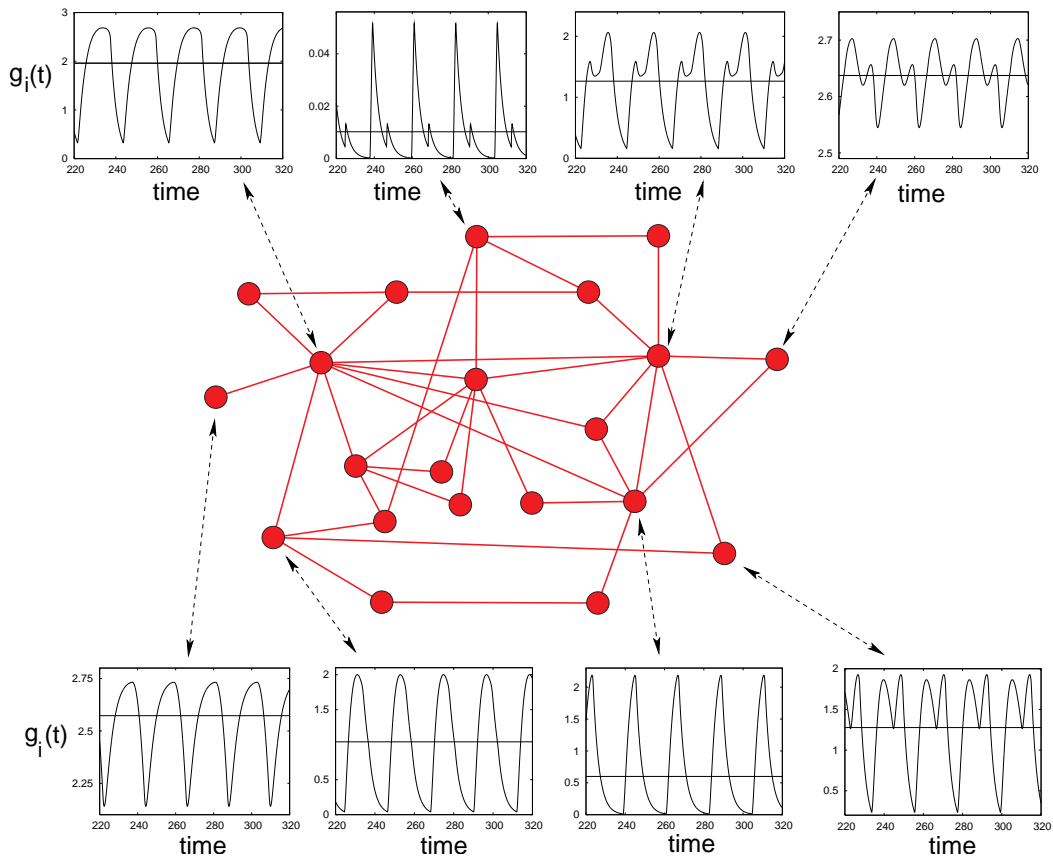}
}
}
\end{tabular}
\caption{
Example of a cluster of $21$ nodes displaying periodic dynamics. 
The insets show the dynamical patterns of each node (see text for the 
interpretation). The maximum Liapunov exponent is $\lambda=-0.00034$ 
and the dynamical parameters are $h=4$ and $p=0.7$. The cluster is 
embedded in a substrate network of $N=50$.
} 
\label{figure3}
\end{figure*}

\begin{figure*}[!htb]
\begin{tabular}{cc}
\centerline{
\resizebox{14.cm}{!}{%
\includegraphics[angle=-0]{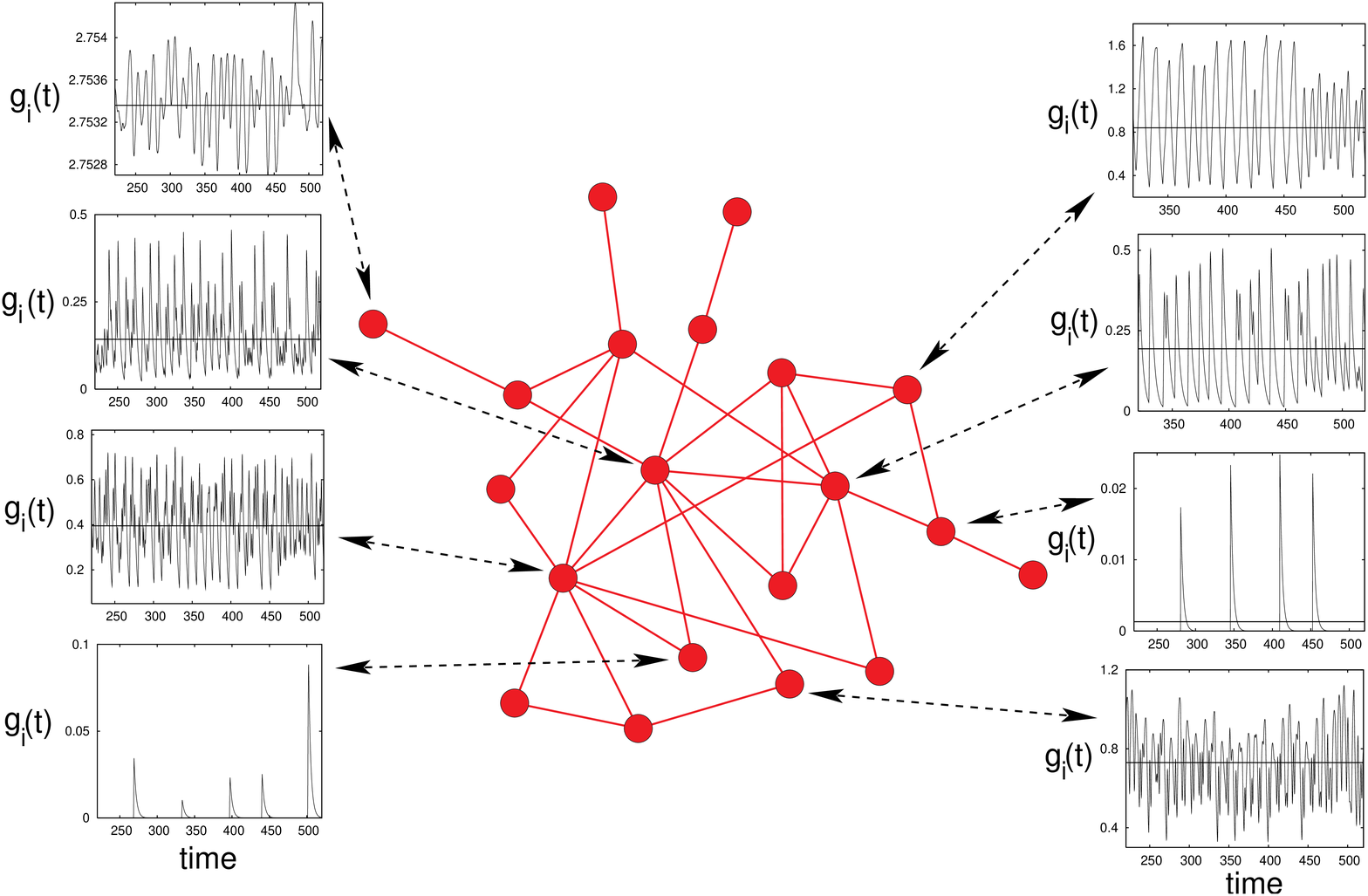}
}
}
\end{tabular}
\caption{Example of a cluster of $19$ nodes displaying chaotic dynamics. 
The insets show the dynamical patterns of each node (see text for the 
interpretation). The maximum Liapunov exponent is $\lambda=0.4716$ 
and the dynamical parameters are $h=4$ and $p=0.7$. The cluster is 
embedded in a substrate network of $N=50$.} 
\label{figure4}
\end{figure*}

From the previous considerations, wether or not the set ${\mathcal D}$
percolates the network realization, leaving out islands of
disconnected activity, is an event that clearly depends on both the
parameter $p$ and the initial conditions. But also the discussion
correctly suggests that fragmentation of the network into subclusters
with independent temporal evolution is a generic (non-zero measure)
feature.  Our numerics, which are extensive in the sense of (both,
network realizations and initial conditions) large sampling,
convincingly corroborate this assertion. Figures\ \ref{figure3} and
\ref{figure4} show two islands of periodic and chaotic activity,
respectively, as well as the temporal evolution of $g_i(t)$ for some
of their constituent's nodes (see the next section for a more detailed
discussion of the figures).

\subsection{Temporal fluctuations of asymptotic solutions.}
\label{subsec:2.2}

The asymptotic dynamics of equation (\ref{eq1}) was studied in
\cite{JGG}. We summarize here the most salient features of the
phase portrait of the collective dynamics.

The presence of inhibitory interactions makes possible the existence
of instabilities in the fixed point solutions ({\em i.e.} states of
constant activities, $g_i(t) = g^*_i$, let us say chemostasis regime)
of evolution equations (\ref{eq1}). Using linear stability analysis
techniques, these "typical" instabilities are characterized as Hopf
bifurcations (either direct or often inverse), where attractors of
exactly periodic collective activities, $g_i(t) = g_i(t+T)$, are born
out from the unstable fixed points. The inverse period (frequency)
$\omega =1/T$ of a periodic attractor changes with parameter and is
naturally dependent on each specific island realization. A sampling
over different initial conditions and network realizations shows that
$\omega$ is smoothly peak- distributed around a value of order unity
(the characteristic time scale of activity decay in the absence of
interactions) with a decaying queue slightly biased to higher
frequency values \cite{JGG}.

\begin{figure}[!tb]
\begin{tabular}{cc}
\centerline{
\resizebox{7.cm}{!}{%
\includegraphics[angle=-0]{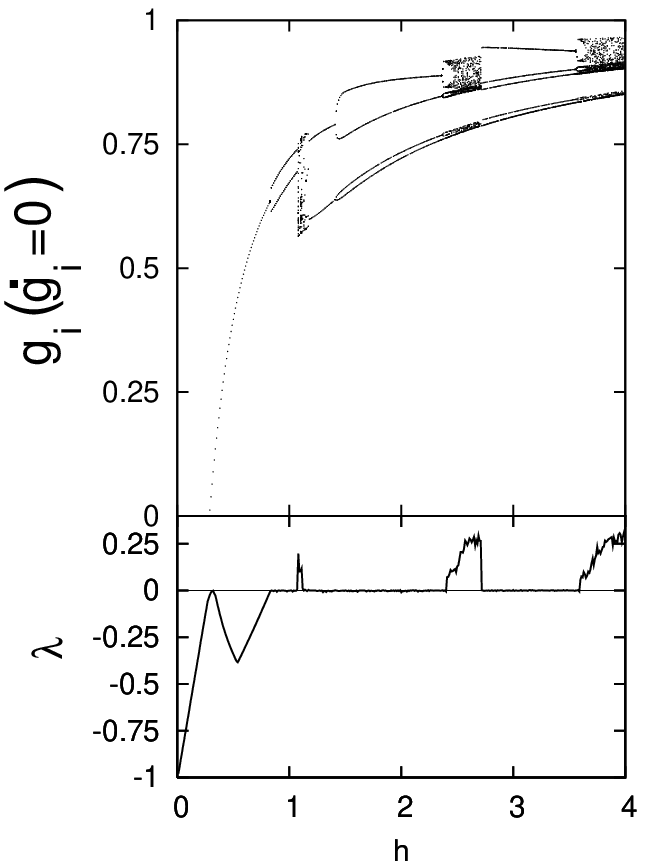}
}
}
\end{tabular}
\caption{Example of bifurcation diagram (island size: 14; $N=60$;
$p=0.8$). One can appreciate an inverse Hopf bifurcation and
several (direct and inverse) period doubling bifurcation cascades.
The maximum Lyapunov exponent $\lambda$ is plotted in the lower
part.} \label{figure5}
\end{figure}

One easily observes that these periodic attractors, in turn,
typically experience also period doubling instabilities, and
through the well-known universal scenario of (successive) period
doubling bifurcation cascade, the onset of chaotic attractors
takes place in the phase portrait of the network dynamics. To help
visualization of the generic types of asymptotic network dynamics,
we represent in figure \ref{figure5} the bifurcation diagram for a
typical attractor. At different values of the (Michaelis-Menten)
parameter $h$, and constant values of $\alpha=3, p=0.7$, we plot
the activity of an individual node at the instant when its time
derivative vanishes. Thus, a single branch in the figure indicates
stationary activity, two branches indicate a periodic attractor,
etc. We also plot in figure \ref{figure5} the largest Lyapunov
exponent $\lambda$ on the attractor, so to allow discerning
between chaotic (positive $\lambda$) and eventual regular
quasiperiodic evolutions ($\lambda = 0$).

\begin{figure}[!htb]
\begin{tabular}{cc}
\centerline{
\resizebox{7.cm}{!}{%
\includegraphics[angle=-0]{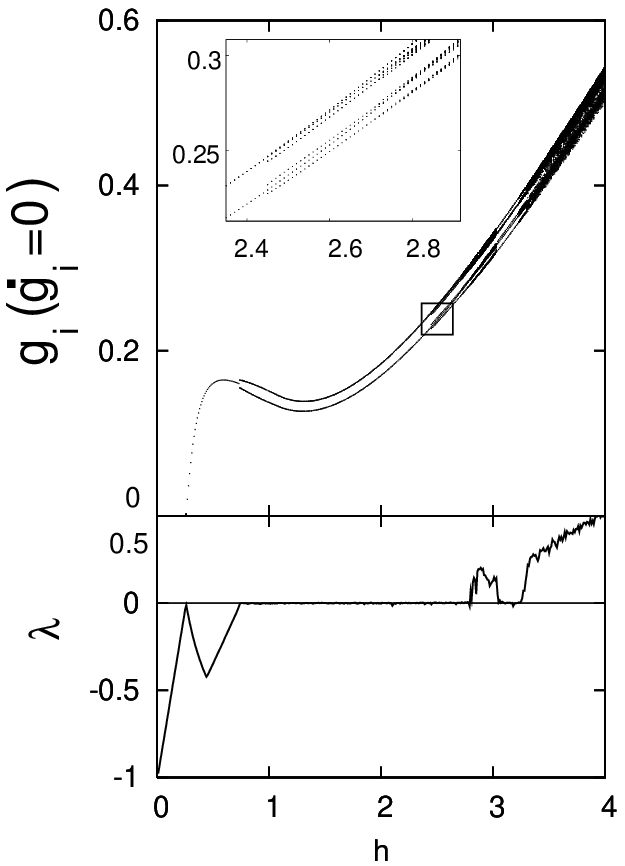}
}
}
\end{tabular}
\caption{Example of bifurcation diagram (island size: 12; $N=60$;
$p=0.8$) showing (see inset) a period tripling bifurcation. The
maximum Lyapunov exponent $\lambda$ is plotted in the lower part.}
\label{figure6}
\end{figure}

A similar bifurcation diagram for a different network realization
is shown in figure \ref{figure6}, where one can appreciate (see
inset) a commonly found bifurcation (though it appears much less
often than period doubling), namely period tripling bifurcation.
Its characterization will be made below in the next subsection
where the Floquet analysis of periodic attractors is presented.

To illustrate the aspect of typical periodic fluctuations we show
in figure \ref{figure3} some examples of the temporal activity
$g_i(t)$ of different nodes inside an island of syncronised
activity from a representative system. Note that the abundance of
out of phase oscillations of neighbours activity is a natural
consequence of inhibitory interactions. Horizontal lines in insets
indicate the average level $\bar{g}_i$ of node activity. We see
that in some of the island nodes the amplitude of the oscillation
is small compared to $\bar{g}_i$ (see {\em e.g.} top rightmost and
bottom leftmost insets); while in others they are of comparable
size, even to the point that lowest levels of activity can reach a
null value, before activity is triggered again after inhibiting
neighbors activity decreases enough.

An analogous visualization of chaotic temporal fluctuations of the
activities in a cluster is shown in figure \ref{figure4}. Here
again we see nodes ({\em e.g.} top left inset) where the size of
activity fluctuations is less than 1 per cent of the average level
$\bar{g}_i$. Most remarkable, there are nodes (like the one in
bottom left inset) which remain inactive most of the time
intermittently experiencing spikes of short duration activity.
This amazing variability of individual node temporal activity on
the chaotic attractors is a generic feature of the network
dynamics. The existence of spike behaviour of individual nodes
activity suggests correctly that eventual variations of parameters
like $h$ may lead to permanent inactivity of some particular
nodes, so providing a straightaway decreasing of the dynamical
cluster size or, the other way around, the activation of inactive
nodes in the frontier.

It is important to note that, for a fixed set of parameter values
and a given network realization, there are generally several
different attractors coexisting in the phase space portrait of the
network dynamics, each one having its own basin (of attraction) of
initial conditions. Multi-stability appears as a generic
consequence of the excitatory/inhibitory interplay. Importantly
also, there can be very many unstable periodic trajectories (often
entangled) flowing in between basins of attractions. The
excitatory/inhibitory competition is also responsible for the appearance
of temporally complex (positive Lyapunov exponent) aperiodic
evolutions, associated to the bifurcation cascade scenario. As we will 
show in section \ref{subsec:3.1} the manifestation of fluctuating (either 
periodic or chaotic) temporal behaviours takes importance when inhibitory 
links predominate, though not too much, over excitatory ones.

\subsection{Floquet analysis of the periodic attractors.}
\label{subsec:2.3}

As shown in the bifurcation diagrams of figures \ref{figure5} and
\ref{figure6}, periodic solutions of the network dynamics often
become unstable under variations of the model parameters. In order
to characterize these instabilities in a precise manner, one may
perform the linear stability analysis of the periodic orbits (see,
{e.g.} \cite{Seydel}) near the bifurcation points.

For this we consider small perturbations of the dynamical
variables, $\delta\vec{g}(t_{0})=\left\{\delta
g_{i}(t_{0})\right\}$, and compute their evolution over the period
$T$ of the periodic orbit. The evolution of these small
perturbations (vectors in tangent space) follows the (linear)
dynamics obtained by linearizing equation (\ref{eq1}) around the
periodic orbit
$\left\{\hat{g}_{i}(t)\right\}=\left\{\hat{g}_{i}(t+T)\right\}$,
{\em i.e.},
\begin{equation}
\frac{d\delta\vec{g}(t)}{dt}=-\delta\vec{g}+ \alpha\cdot{\mathcal
A}\delta\vec{g}\;, \label{eq3}
\end{equation}
where the matrix ${\mathcal A}$ is obtained
as
\begin{equation}
{\mathcal A}_{i,j}=\frac{\Theta[\sum_{k}W_{i,k}g_{k}]
}{(1+h^{-1}\Phi[\sum_{k}W_{i,k}g_{k}])^2}\cdot W_{i,j} \label{eq4}
\end{equation}
and $\Theta[x]$ denotes the (Heaviside) step function. Note that the above 
equation is only valid when the sum of the inputs (activatories and 
inhibitories) which receives a node from its neighbours is nonzero. 
Hence, the Floquet analysis is performed for each dynamical cluster found 
and not for the whole network.  

\begin{figure}[!tb]
\begin{tabular}{cc}
\centerline{
\resizebox{8.cm}{!}{%
\includegraphics[angle=-0]{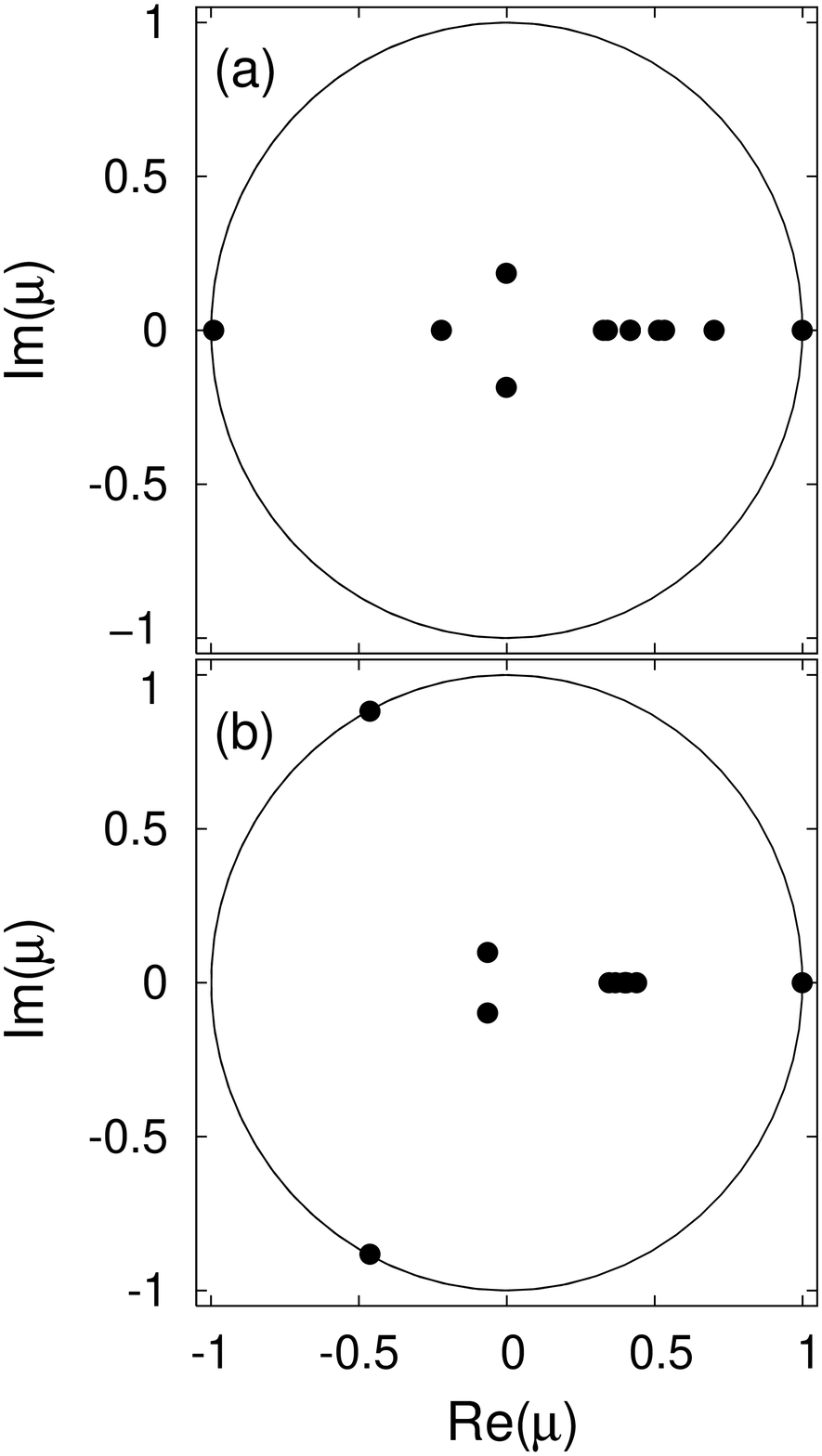}
}
}
\end{tabular}
\caption{Floquet spectra. {\bf (a)} Period doubling bifurcation in
an island of size 14 at $h=1.57$. {\bf (b)} Naimark-Sacker
bifurcation (at rational Floquet angle $\theta=\pm2\pi/3$) in an
island of size 12 at h=2.44 (the same as used in the diagram of
figure \ref{figure6}). For both $N=60$ and $p=0.8$.}
\label{figure7}
\end{figure}

The so-called Floquet (or monodromy) matrix ${\mathcal F}$ of the
periodic solution $\left\{\hat{g}_{i}(t)\right\}$ is defined as
the linear operator in tangent space that maps the initial
perturbation at $t_0$, $\delta\vec{g}(t_{0})$, onto the
perturbation at $t_0+T$
\begin{equation}
\delta\vec{g}(t_{0}+T)={\mathcal F}\delta\vec{g}(t_{0})
\end{equation}

\begin{figure*}[!htb]
\begin{tabular}{cc}
\centerline{
\resizebox{15.cm}{!}{%
\includegraphics[angle=-0]{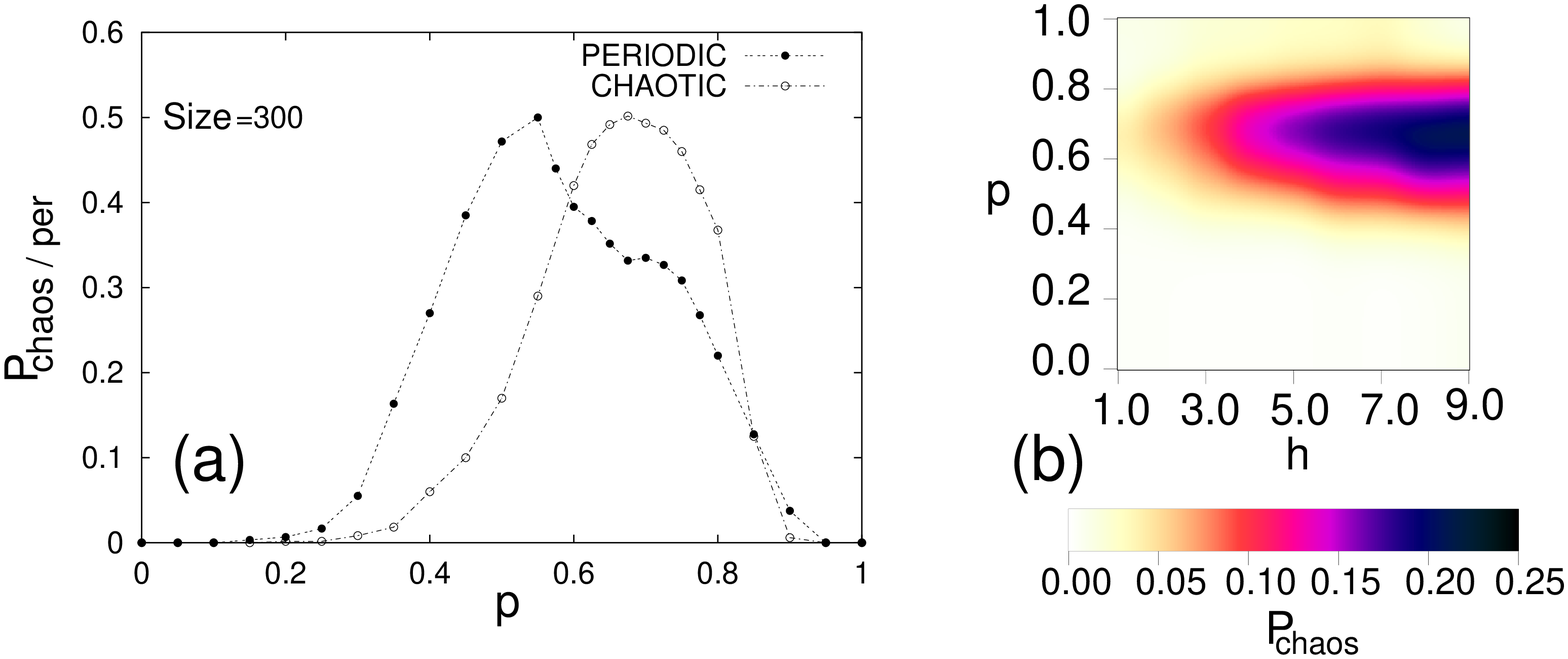}
}
}
\end{tabular}
\caption{(Color online){\bf (a)} Probability, $P_{chaos}$ ($P_{per}$), that the system
    evolves to a chaotic (periodic) regime as a function of the
    probability of inhibitory interactions, $p$, for $h=4$ and
    $N=300$. {\bf (b)} Phase diagram in the $(p,h)$-parameter space of 
    the chaotic dynamics of the system. Color code indicate the values of 
    $P_{chaos}$ ($N=300$).}
\label{figure8}
\end{figure*}

The Floquet matrix ${\mathcal F}$ is obtained by numerical integration of the
linearized equation (\ref{eq3}) over a period $T$ for a basis of
initial conditions in the tangent space. The spectrum of eigenvalues
of this matrix provides the information on the linear stability of the
periodic solution. Note that because ${\mathcal F}$ is a real matrix,
if a Floquet eigenvalue $\mu$ is a complex number, then its complex
conjugate $\bar{\mu}$ also belongs to the Floquet spectrum. Also,
because solutions of autonomous differential equations can be shifted
in the time $t$ direction, their Floquet matrix always has unity as an
eigenvalue, say $\mu^1=1$, with associated eigenvector
$\{\dot{\hat{g}}_{i}(t_0)\}$. The solution is linearly
stable if all the other eigenvalues $\mu^j=|\mu^j|\exp(i\theta^j)$ are
in the interior of the unit circle of the complex plane, {\em i.e.}
$|\mu^j|<1$ for $j \neq 1$. A periodic solution becomes unstable when
a Floquet eigenvalue (or a pair of complex conjugate eigenvalues)
crosses the unit circle. The associated Floquet eigenvector indicates
the direction in tangent space where perturbations will grow
exponentially away from the solution.

In figure \ref{figure7}(a) we plot the Floquet spectrum of a periodic
attractor at a period doubling bifurcation. As seen in the figure, a
Floquet eigenvalue crosses the unit circle at $-1$.  In figure
\ref{figure7}(b) we plot the Floquet spectrum of the periodic
attractor of figure \ref{figure6} at $h=2.44$, where the inset
suggested that a period tripling bifurcation may occur. We see a
complex conjugate pair of Floquet eigenvalues exiting the unit circle
at angles $\theta=\pm2\pi/3$. In general, for generic irrational
values of $\theta/\pi$ this type of bifurcation (called Naimark-Sacker
or generalized Hopf bifurcation) gives rise to a quasiperiodic
attractor whose trajectories fill densely a two-frequency
torus. However, as a generic feature of our model, the two frequencies
of the new attractor are in a commensurate ratio ($2:3$), so that the
new stable trajectory has a period of 3T.

In terms of how often different types of bifurcation occur in the
network dynamics, as inferred from our (non-exhaustive, but
significant at the scales considered) sampling of initial conditions and
network realizations, one may say that period doubling cascades
and, less often, commensurate Naimark-Sacker bifurcations have
been generically found by varying the Michaelis-Menten parameter
$h$. But, besides the formal characterization of the dynamical
instabilities observed, the Floquet analysis allows also to give
an answer on a more general question, namely how temporal
instabilities correlate with networking connectivity
characteristics. Are there characteristic features discernible in
the structure of instabilities? This point will be discussed
further below in the next subsection.

\section{Statistical characterization of dynamical regimes and islands structure.}
\label{sec:3}

As noted before, the dynamics of the system is determined by only two
parameters, $h$ and $p$. The behaviour of the system described by
equation (\ref{eq1}) on the underlying network is very rich and one
can have steady, periodic or chaotic states as well as
fragmentation. In this section, we analyze in more details the
system's phase diagram as well as how the dynamical regimes couple to
the local structural properties of the underlying network and
dynamical islands.


\subsection{Density distribution functions of dynamical regimes.}
\label{subsec:3.1}

In figure \ref{figure8}, we have represented the probability,
$P_{chaos}$, of ending up in a chaotic regime as a function of $p$ for
a network of $N=300$ nodes and $h=4$. This probability is given by the
fraction of the total number of realizations (typically $10^3$
different initial conditions over different network realizations for
each value of $p$ and $h$ were used) in which at least one chaotic
dynamics is observed. The figure also shows the corresponding
probability, $P_{per}$, for periodic orbits. As figure 
\ref{figure8}(a) clearly shows, there is a threshold value
$p=p_{th}$ beyond which the network dynamics is not robust under
variations of the initial values of the $g_i$'s. For values of $p$
above $p_{th}\approx 0.25(5)$, two randomly chosen initial conditions
can lead the system to disparate asymptotic regimes. Besides, the size
of the system affects the value of $P_{chaos}$, but the onset $-$and
the end$-$ of the chaotic phase seems to be $N$ independent
\cite{JGG}.

\begin{figure}[!htb]
\begin{tabular}{cc}
\centerline{
\resizebox{8.cm}{!}{%
\includegraphics[angle=-0]{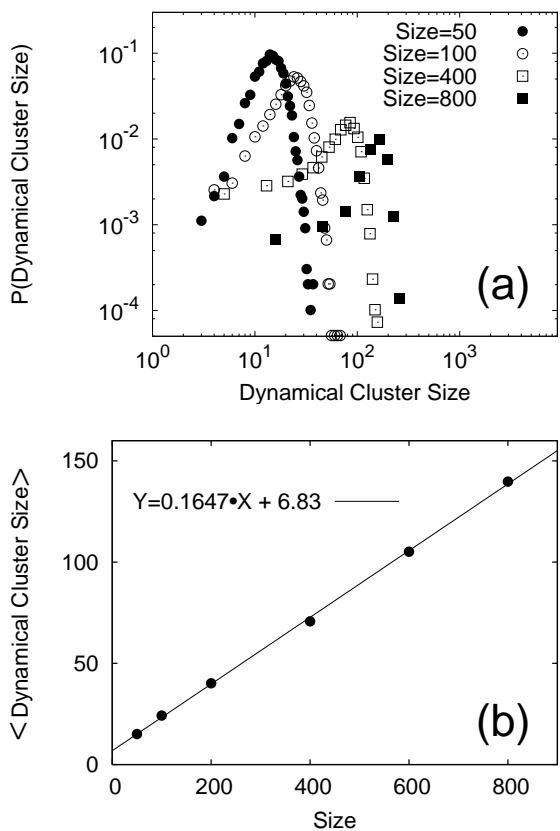}
}
}
\end{tabular}
\caption{{\bf (a)} Probability that a connected cluster of nodes
displaying either chaotic or periodic behavior has a given size (in
number of nodes forming the cluster). {\bf (b)} Scaling of the mean
cluster size with $N$. The parameters have been set to $h=4$ and
$p=0.7$.} \label{figure9}
\end{figure}

Moreover, figure \ref{figure8}(a) constitutes a quantitative
illustration of how the prevalence of fluctuating asymptotic regimes
over chemo-stasis ones depends on the model parameter $p$. The sum of
both functions, $P_{per}(p) + P_{chaos}(p)$, gives the probability
that the asymptotic state shows temporal variations of the activity
vector (either regular or chaotic) as a function of $p$. These results
give that in the range of values $0.5 \leq p \leq 0.8$ regimes of
temporal fluctuations occur more often than constant activity
regimes. This measure is maximized by values around $p \simeq 2/3$
and, quite naturally, it increases with the value of the
Michaelis-Menten parameter $h$, {\em i.e.} the slope at the origin of
the saturated response function (see figure \ref{figure1}). Note that
even larger values of $p$ means overabundance of inhibitory
interactions, which leads to the predominance of the asymptotic rest
state, while smaller values of $p$ favour chemostatic equilibria.

The quantities $P_{chaos}$ and $p_{th}$ depend on $h$. As we move to
larger values of $h$, the strength of the interactions increases and
hence it is expected that slight perturbations produce a behavior in
which the fraction of nodes whose dynamical patterns are easily
disturbed grows. This is indeed the case, as illustrated in figure 
\ref{figure8}(b). The color-coded figure shows that as $h$ is
increased, the probability of having a chaotic phase grows, and that
the onset of such chaotic patterns shifts to smaller values of
$p$. This drift of $p_{th}$ is however bounded. For small enough
values of $p$ (even for very large $h$), most of the elements activate
each other ($W_{ij}=1$ for a large fraction of pairs ${ij}$ and
${ji}$) and hence the resulting dynamics is steady. In other words,
the onset of chaotic regimes is always located at a nonzero value of
$p_{th}$ (the same applies to the right (decaying) part of
$P_{ch}(p)$, but in this case the activity falls down to
zero). 

\begin{figure}[!tbh]
\begin{tabular}{cc}
\centerline{
\resizebox{8.cm}{!}{%
\includegraphics[angle=-0]{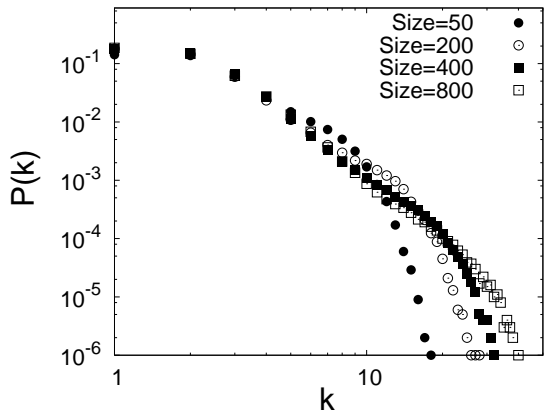}
}
}
\end{tabular}
\caption{Probability that a node belonging to a dynamical island
  interacts with $k$ other nodes of the island. 
  Parameters were set to $h=4$ and $p=0.7$.} \label{figure10}
\end{figure}

\begin{figure}[!tbh]
\begin{tabular}{cc}
\centerline{
\resizebox{8.cm}{!}{%
\includegraphics[angle=-90]{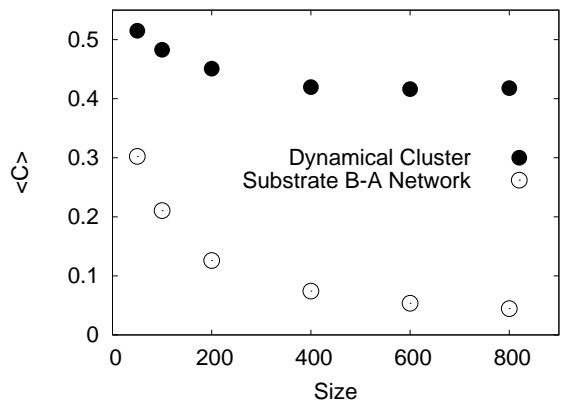}
}
}
\end{tabular}
\caption{Average clustering coefficient $\langle C \rangle$ as a
  function of the network size for the BA original network and the
  dynamical cluster. Note that while $\langle C \rangle$ in the BA
  network continuously decreases, for the dynamical island it
  saturates. See more details in the text. The results have been
  obtained using $h=4$ and $p=0.7$.} \label{figure11}
\end{figure}


\subsection{Dynamical island structure}
\label{subsec:3.2}

We next focus on the topological characterization of islands of
dynamical units. We first analyze how the cluster size distribution of
islands of nodes displaying either periodic or chaotic activity scales
with the system size. Figure \ref{figure9}(a) represents the
probability that an island has a given size for several networks made
up of a number of nodes ranging from 50 to 800. Clearly, the size
distribution shows an average value that changes as $N$ grows. A
closer look at the figure (see figure \ref{figure9}(b)) reveals that the
mean cluster size scales with $N$ and that about $17\%$ of the nodes,
in average, exhibits nonzero activity. This confirms what we have
discussed in section \ref{fragmentation} about the measures of the sets
${\mathcal D}^{*}$ and ${\mathcal D}$, namely, that the fragmentation
of the network into islands of independent dynamics appears as one of
the most characteristic features of the model.

\begin{figure*}[!htb]
\begin{tabular}{cc}
\centerline{
\resizebox{17.cm}{!}{%
\includegraphics[angle=-0]{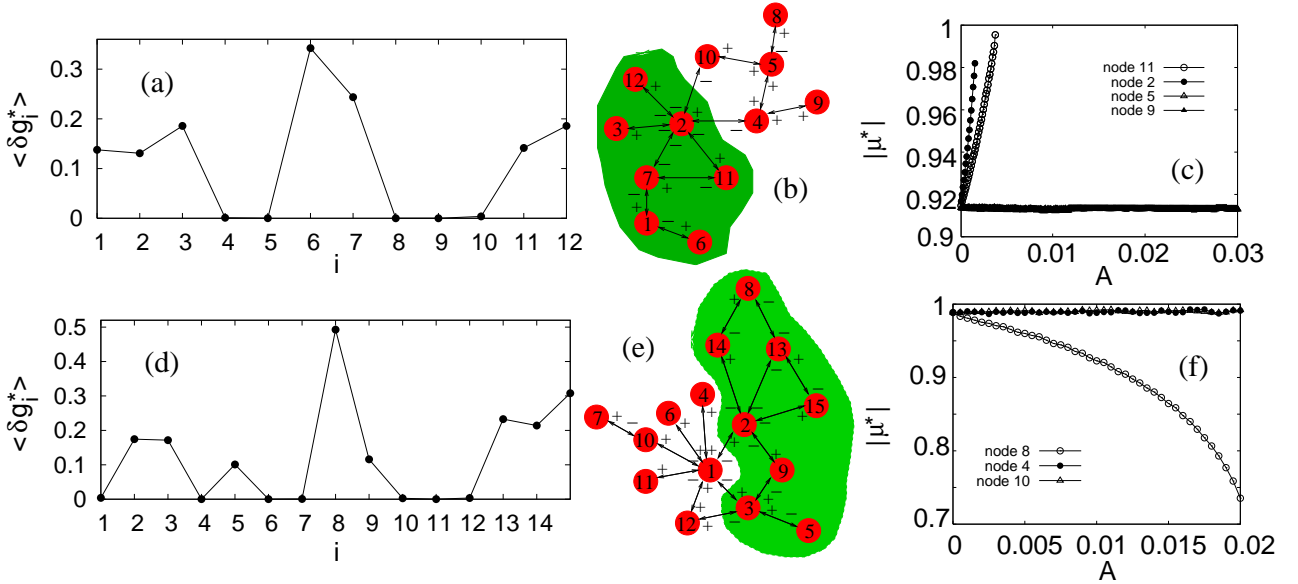}
}
}
\end{tabular}
\caption{(Color online) {\bf (a)} and {\bf (d)} show the components of
the vector $\langle \vec{{\delta g}}^{\star}\rangle$ (see text) for
two dynamical islands at the critical point of a Naimark-Sacker
bifurcation at $h=2.44$ {\bf (a)} and a period doubling one at
$h=2.629$ {\bf (d)}. {\bf (b)} and {\bf (e)} show the distribution
(green region) of the nodes with non null component of $\langle
\vec{{\delta g}}^{\star}\rangle$ in {\bf (a)} and {\bf (d)}
respectively. Finally, {\bf (c)} and {\bf (f)} show the evolution of
the Floquet eigenvalue $\mu^{\star}$ as a function of the forcing
amplitude $A$ applied to different nodes of the dynamical islands {\bf
(b)} and {\bf (e)} respectively. For both islands the susbstrate
network was of $N=60$ nodes whith a fraction of $80\%$ of inhibitory
interactions ($p=0.8$).}
\label{fig:Comp}
\end{figure*}

Another interesting aspect is the elucidation of how the topological
properties of the islands correlate with those of the underlying
(original) network. To this end, we have further scrutinized the
structure of the clusters and measured two topological quantities of
interest. Figure\ \ref{figure10} shows the degree distribution of
nodes belonging to dynamical islands for several system sizes. This
property can be regarded as a global one and indicates that within the
islands, the probability that a node has $k$ links also follows a
power-law, though with a more pronounced cut-off and a (slightly)
different value for the exponent $\gamma$. More striking is the result
depicted in figure \ref{figure11}, where the average clustering
coefficient $\langle C \rangle$ of the substrate (original) network
and of the islands is plotted as a function of $N$. While for the BA
network the clustering is vanishing as the network size grows, it
seems that for dynamical islands its value saturates. This is quite
interesting because, on the other hand, the value of the clustering
coefficient is very large and comparable to measures of real systems
where the kind of dynamics explored here applies, for instance,
biological networks \cite{borev}.

That is, the structure of dynamical islands correctly reproduces
several of the most important topological features observed in
biological networks and not captured by current network
models. Namely, the heterogeneous distribution of connections, a high
average clustering and the independence of $\langle C \rangle$ with
respect to the system size. This result points to the conjecture that
several topological properties observed in systems driven by AI
interactions where nodes are themselves (nonlinear) dynamical units
may be biassed by their own dynamics. In other words, what we actually
see is the result of the activity showed up by a smaller ``dynamical''
network whose local topological properties greatly differ from those
of a larger substrate network that we don't ``see'' because many of
its components are simply off. This, in fact, may be the case of
biological systems where structure and dynamics are indissoluble
linked \cite{strogatz,borev}.

\subsection{Structure inside Dynamical Islands}
\label{subsec:3.3}

The above findings on new (dynamically) emergent characteristics of
the islands structure motivates the question of wheter these clusters
have an internal organization or hierarchy among its constituents.  It
is widely known that when one deals with problems where the network
topology (scale-free) is the only degree of complexity of the problem
the answer to this question is usually based on the presence of highly
conected nodes (the hubs). This is the case when linear evolution
equations are studied on top of complex networks like epidemic or
rumour spreading, traffic and communication problems
\cite{mnv04,Echenique2005a,Echenique2005b}, However, our case is not
so simple and the nonlinear excitatory/inhibitory dynamics between the
elements of the network plays a crucial role in determining which
nodes are governing the evolution of the system.  Moreover, the high
clustering found for the dynamical clusters points out that this
leading role is not played by isolated nodes but by small
substructures inside the dynamical islands. This concept is not new,
the problem of finding small relevant substructures inside large
networks, usually called ``{\em motifs}'' \cite{milo2}, has been
studied in different ways in the field of biological networks.

It is indeed very revealing to pay attention to the {\em networked}
structure of the unstable manifold, which is given in the linear
regime of small perturbations by the Floquet unstable
eigenvectors. For this purpose, we look at the behavior of the
components of the dynamical islands when a bifurcation (either period
doubling or Naimark-Sacker type) occurs. In these critical points, it
is possible to get a deeper insight into what is going on in the
dynamical islands $I$  by
looking at the Floquet eigenvector responsible for the bifurcation,
$\vec{\delta g}^{\star}(t_0)= \{\delta g_{i}^{\star}(t_0)\}$,
corresponding to the Floquet eigenvalue which reaches the unit
circle. In particular, integrating equation (\ref{eq3}) for the
initial condition $\vec{\delta g}^{\star}(t_0)$ we can compute the
following vector
\begin{equation}
\vec{\langle {\delta g}^{\star}\rangle}=\left\{\langle {\delta
g}_{i}^{\star}\rangle\right\}= \left\{\frac{1}{T}\int_{0}^{T}{|\delta
g_{i}^{\star}|dt}\right\}\;.
\label{eq6}
\end{equation}

The components of this vector measure, for each node, the average
(over a period $T$ of the old solution) distance of the new solution
after the bifurcation point from the old periodic solution. Note that
a zero component of this vector at a node $k$, means orthogonality of
the single-site perturbations at that node with respect to the
unstable direction in tangent space. In other words, by looking at the
components of the vector (\ref{eq6}) we can identify those nodes that
are more affected by the perturbation that leads the system to
instability. In figures \ref{fig:Comp}(a) and \ref{fig:Comp}(d) we
show this quantity for the same two islands (relatively small, but
still representative) whose Floquet spectra are given in Fig.\
\ref{figure7}, one (a) corresponding to a Naimark-Sacker bifurcation
and the other (d) to a period doubling bifurcation.

As it can be seen from the figures, the vectors $\vec{\langle {\delta
g}^{\star}\rangle}$ have several null components. The structural
profiles reveal, apparently irrespective of the type of instability,
that the set $\mathcal{S}$ of nodes in the island which are alien to
instability (white regions), that is, the set of those nodes $k$ such
that $\langle \delta g_k \rangle =0$, is a non-zero measure set; it is
sometimes even larger than the complementary set (green area)
$\mathcal{U}=I-\mathcal{S}$ of participating nodes on the unstable
eigenvector evolution during a period.  We observe here that the
fragmentation tendency (see discussion on islands of disconnected
dynamics made above) operates also at the level of the tangent space,
in the sense that a binary partition of the island nodes is well
defined at the bifurcation (critical) point. Namely, the instability
introduces a partition of the island $I = \mathcal{U} \oplus
\mathcal{S}$ into (a) the set $\mathcal{U}$ of nodes that do
participate in the instability evolution in the linear regime, and (b)
the complementary set $\mathcal{S}$, of nodes such that single-node
perturbations are orthogonal to the unstable linear manifold.  This
drastic, generic fragmentation of the island of periodic activity at
the linear description level of the bifurcation, is also clearly the
consequence of the AI competition on the network of interactions, and
we have not seen any deviation from this observation in the
computations performed (of which only two cases are illustrated). In
summary, one could say that inside the dynamical islands there
are compact substructures (and not single nodes) governing the
dynamical changes of the whole cluster of nodes.

The behavior described above suggest the following numerical
experiment: we have explored the responses of the different nodes to
an external perturbation when the system is in a periodic state near a
bifurcation point. In particular, we force a node by adding an
aditional term of the form $A\cdot \sin (\omega t)$ (with $\omega=
2\pi/T$ where $T$ the period of the unperturbed system) to its
equation of motion (\ref{eq1}). Then we compute, as a function of the
forcing amplitude $A$, the evolution of the Floquet eigenvalue
$\mu^{\star}$ responsible for the forthcoming bifurcation in the
unperturbed system.  The effects of such a perturbation strongly
depend on whether the perturbed node belongs to the subset of those
identified as leaders ({\em i.e} the ones with non null component in
$\vec{\langle {\delta g}^{\star}\rangle}$). The results obtained for
the two dynamical islands aforediscussed are shown in figures
\ref{fig:Comp}(c) and \ref{fig:Comp}(f). When the nodes inside the
green area are perturbed the Floquet eigenvalue $\mu^{\star}$
significantly deviate (either increase or decrease, we have not been
able to elucidate when a given change is expected) from the values of
the unperturbed system. On the other hand, the perturbation of the
nodes located outside the green region does not imply any change to
linear stability of the whole system. These results illustrate the
relevant role played by the substructures found by the computation of
$\vec{\langle {\delta g}^{\star}\rangle}$.

\section{Concluding remarks.}
\label{sec:4}

In this paper, we have analyzed the interplay between complex
topologies and activatory-inhibitory interactions driven by a
saturated response dynamics of the Michaelis-Menten type. The dynamics
of the system is very rich and exhibits steady, periodic and chaotic
regimes that in turn lead to the fragmentation of the original
substrate network into a smaller cluster of dynamically active
nodes. We have fully characterized these states by means of the
Lyapunov exponent and the Floquet analyses as well as the topological
features of active islands. The reach behavioral repertoire observed
is thus a consequence of the entangled complexity of the system
temporal behavior and the heterogeneous structure of the underlying
network. 

The emerging dynamics characterized in this work could plausible
describe at least two relevant scenarios in biological systems. On one
hand, the dynamics expressed in Eq.\ (\ref{eq1}) has been proposed as
a way to characterize theoretically the individual dynamics of gene
expression \cite{sole}. In fact, Eq.\ (\ref{eq1}) is the
generalization of the successful Random Boolean models
\cite{kauffman,oosawa} widely used to model gene expression. In this
context, two nodes at the ends of a link are considered to be
transcriptional units which include a regulatory gene. One of these
end-nodes can be thought of as being the source of an interaction (the
output of a transcriptional unit). The second node represents the
target binding site and at the same time the input of a second
transcriptional unit. By studying simplified models as the one
implemented here $-$ the intrinsic complexity of the problem does not
allow for a complete and detailed description of real gene dynamics
$-$, one can infer the region of the parameter space (i.e. $(p,h)$)
that better describes gene networks. The latter seems possible due to
latest developments in microarray technologies, biocomputational
tools, and data collection software.

A second scenario where the results obtained apply is reaction
kinetics in metabolic networks. In metabolic systems, a very rich
behavioral repertoire is well documented \cite{murray}, as for
instance, the oscillations observed in the concentration of certain
chemicals in biochemical reactions such as glycolysis. The system of
differential equations, Eqs.\ (\ref{eq1}), represents one of the most
basic biochemical reactions, where substrates and enzymes are involved
in a reaction that produces a given product. In this context, there
are several important issues as how fast the equilibrium is reached,
how the concentration of substrates and enzymes compare, etc. Besides,
it is known that in a large number of situations, some of the enzymes
involved show periodic increments in their activity during division,
and these reflect periodic changes in the rate of enzyme synthesis.
This is achieved by regulatory mechanisms that necessarily require
some kind of feedback control as that emerging in our model. The
interesting point here is that the real topological features of the
underlying metabolic network \cite{metab} have not been taken into
account in studies performed so far. As this work shows, they have
important bearings in the correlation between structure and the
observed dynamics.

Finally, on more general theoretical grounds, we anticipate several
features of interest such as the fragmentation of the original network
according to the dynamical states of the nodes, multistability and
different routes to chaotic behavior within the same system. The first
of these points is particularly relevant since it may indicate that in
networks of dynamical units, the topology observed can be the result
of a given network state hiding a larger substrate whose topological
properties are completely different at a local level. Of particular
interest is also the result gathered in the last part of the work,
namely, the existence of an additional substructure inside dynamical
islands determined by the different responses of nodes to external
perturbations. This points to the central issue in many biological
processes of what subset of nodes are the most important in order to
sustain (or break) the system's robust functioning. In summary, the
characterization of models where nonlinearity and spatial complexity
coexist yields new results missed when only one of these ingredients
is present and opens the path to a better comprehension of biological
processes and the dynamics of networks of nonlinear dynamical units.

\begin{acknowledgements}

The authors thank S.\ Boccaletti and S. Manrubia for useful
suggestions and discussions. J.\ G-G\ acknowledges financial support
of the MECyD through a FPU grant. Y.\ M.\ is supported by MEC through
the Ram\'on y Cajal program. This work has been partially supported by
the Spanish DGICYT projects BFM2002-00113, BFM2002-01798 and
FIS2004-05073-C04-01; and by DGA (grupo consolidado FENOL).

\end{acknowledgements}

\end{document}